\documentclass[manuscript,screen]{acmart}

\usepackage{amsmath,amsfonts,amsthm}
\usepackage{algorithm,algorithmic}
\usepackage{xspace}
\usepackage{booktabs}
\usepackage{graphicx}
\usepackage{textcomp}
\usepackage{xcolor}
\usepackage{listings}
\usepackage{multirow}
\usepackage{relsize}
\usepackage{pifont}
\usepackage{hyperref}
\usepackage{subfig,paralist}
\usepackage{pifont}
\usepackage{enumitem}
\usepackage{url}
\usepackage{booktabs}
\usepackage{tabularx}
\usepackage{multirow,array}
\usepackage{threeparttable}
\usepackage{bm}

\usepackage{tikz}

\def\BibTeX{{\rm B\kern-.05em{\sc i\kern-.025em b}\kern-.08em
    T\kern-.1667em\lower.7ex\hbox{E}\kern-.125emX}}

\newtheorem*{theorem*}{Theorem}

\usepackage{tcolorbox}
\definecolor{mycolor}{rgb}{0.122, 0.435, 0.698}
\definecolor{gray1}{gray}{0.3}

\definecolor{codegreen}{rgb}{0,0.6,0}
\definecolor{codegray}{rgb}{0.5,0.5,0.5}
\definecolor{codepurple}{rgb}{0.58,0,0.82}
\definecolor{backcolour}{rgb}{0.95,0.95,0.92}
\lstdefinestyle{mystyle}{
    commentstyle=\color{codegreen},
    keywordstyle=\color{magenta},
    numberstyle=\tiny\color{codegray},
    stringstyle=\color{codepurple},
    basicstyle=\tiny\ttfamily,
    breakatwhitespace=false,
    breaklines=true,
    captionpos=b,
    keepspaces=true,
    numbers=left,
    numbersep=5pt,
    showspaces=false,
    showstringspaces=false,
    showtabs=false,
    tabsize=2,
    columns=fixed
}
\newcommand{\result}[1]{%
\begin{tcolorbox}[colframe=mycolor,boxrule=0.5pt,arc=4pt,
      left=6pt,right=6pt,top=6pt,bottom=6pt,boxsep=0pt,width=\columnwidth]%
      {\emph{#1}}
\end{tcolorbox}%
}

\definecolor{darkgreen}{rgb}{0.0, 0.5, 0.0}
\definecolor{darkred}{rgb}{0.82, 0.1, 0.26}

\newcommand{\todo}[1]{\noindent\textcolor{red}{TODO:\xspace #1}}

\newcommand{\changed}[1]{\textcolor{blue}{#1\xspace}}
\newcommand{\changedminor}[1]{\textcolor{purple}{#1\xspace}}

\renewcommand{\todo}[1]{}

\renewcommand{\changed}[1]{#1\xspace}
\renewcommand{\changedminor}[1]{#1\xspace}

\begin{document}
\title{\changed{Fuzzing: On Benchmarking Outcome as a Function of Benchmark Properties}}

\author{Dylan Wolff}
\email{wolffd@comp.nus.edu.sg}
\affiliation{%
\institution{National University of Singapore}
\country{Singapore}}

\author{Marcel B{\"o}hme}
\email{marcel.boehme@mpi-sp.org}
\affiliation{%
\institution{Max Planck Institute for Security and Privacy}
\country{Germany}}

\author{Abhik Roychoudhury}
\email{abhik@comp.nus.edu.sg}
\affiliation{%
\institution{National University of Singapore}
\country{Singapore}}

\renewcommand{\shortauthors}{}

\begin{abstract}\changed{
In a typical experimental design in fuzzing, we would run two or more fuzzers on an appropriate set of benchmark programs plus seed corpora and consider their ranking in terms of code coverage or bugs found as outcome.
However, the specific characteristics of the benchmark setup clearly can have some impact on the benchmark outcome.
If the programs were larger, or these initial seeds were chosen differently, the same fuzzers may be ranked differently; the benchmark outcome would change. 
In this paper, we explore two methodologies to \emph{quantify the impact of the specific properties on the benchmarking outcome}.
This allows us to report the benchmarking outcome counter-factually, e.g., ``If the benchmark had larger programs, this fuzzer would outperform all others''.
%
Our first methodology is the \emph{controlled experiment} to identify a causal relationship between a single property in isolation and the benchmarking outcome.
The controlled experiment requires manually altering the fuzzer or system under test to vary that property while holding all other variables constant.
By repeating this controlled experiment for multiple fuzzer implementations, we can gain detailed insights to the different effects this property has on various fuzzers.
However, due to the large number of properties and the difficulty of realistically manipulating one property exactly, control may not always be practical or possible.
Hence, our second methodology is \emph{randomization} and non-parametric regression to identify the strength of the relationship between arbitrary benchmark properties (i.e., covariates) and outcome.}
Together, these two fundamental aspects of experimental design, \emph{control} and \emph{randomization}, can provide a comprehensive picture of the impact of various properties of the current benchmark on the fuzzer ranking. 
These analyses can be used to guide fuzzer developers towards areas of improvement in their tools and allow researchers to make more nuanced claims about fuzzer effectiveness.
We instantiate each approach on a subset of properties suspected of impacting the relative effectiveness of fuzzers and quantify the effects of these properties on the evaluation outcome.
In doing so, we identify multiple properties, such as the coverage of the initial seed-corpus and the program execution speed, which can have statistically significant effect on the \emph{relative} effectiveness of fuzzers.
\end{abstract}


\maketitle

\section{Introduction}
\emph{Fuzzing} \cite{bohme2020fuzzing} is a well-known automated software testing method for finding security flaws by generating invalid or unexpected inputs.
In particular, greybox fuzzers, which leverage light-weight instrumentation feedback to guide test input generation, have emerged as one of the most successful automatic bug finding approaches in practice \cite{ossfuzz2}.
Fuzzing has also emerged as an important research topic, \changed{with over 50 fuzzing papers published in the ``Big Four'' academic computer security conferences in 2024 alone \changedminor{(i.e., CCS, NDSS, S\&P, USENIX Security)}!}

\vspace{0.1cm}
\noindent
\emph{Yet, which fuzzer performs best and when?}

Recently, the fuzzing community has identified \emph{sound fuzzer evaluation} as a \mbox{Top-3} most important research challenge~\cite{bohme2020fuzzing}. 
To demonstrate improvement over the state-of-the-art, many fuzzing related benchmarks have been introduced. For instance, the MAGMA benchmark \cite{hazimeh2020magma} provides a set of 138 real bugs in 9 programs. ProFuzzbench \cite{profuzzbench} offers access to 11 protocol implementations for network-enabled fuzzers. The FuzzBench benchmark \cite{metzman2021fuzzbench} offers access to over 650 open-source programs \changed{via an integration with the OSS-Fuzz project~\cite{ossfuzz}}. FuzzBench is officially developed by Google demonstrating a substantial practical interest in sound fuzzer evaluation.
%
Using these benchmarks, two or more fuzzers are compared by ranking them in terms of their performance (e.g., coverage achieved or \#bugs found)  \cite{arcuri2014hitchhiker,klees2018evaluating,benchmarking}.
To ensure that the observed performance differences are not due to random effects, tool developers are encouraged to repeat the experiments at least twenty times and measure {effect size} and {statistical significance} \cite{klees2018evaluating, arcuri2011practical}.
To ensure that the observed performance differences can be attributed precisely to the proposed improvements, tool developers are encouraged to compare the prototype to the baseline which was extended to implement the improvement.
To ensure that the observed performance differences are general, benchmark designers attempt to select a sample of subjects that are representative of the population of systems. 
Within this sample, there can be wide variations in benchmark properties by virtue of it being composed of disparate representative programs.
 
We observe that the benchmark outcomes depend critically on the specific properties of the selected benchmark. 
\emph{On the average}, most fuzzers perform similarly while for each \emph{specific} program there are often clear winners. For instance, in a recent FuzzBench experiment involving 23 programs and 11 fuzzers, we can see that the average ranking for the
majority of fuzzers is $5.5\pm 1$\footnote{\url{https://www.fuzzbench.com/reports/2022-04-19/index.html}} -- most fuzzers rank approximately the same. 
Looking only at these overall rankings across all benchmarks, it is not apparent that e.g. the ranking of AFL++ improves on larger programs.
We call this evaluation methodology \emph{atomistic}, because it does not account for the effects of benchmark properties.
In other words, the final outcome of such an \emph{atomistic} evaluation is \emph{specific} to the current choice of benchmark and provides no insights into the \emph{conditions} under which one fuzzer performs better than another.
We are not the first to make this observation; it is well known that some additional variables, i.e., \emph{covariates}, can have a different relative impact on fuzzer effectiveness, and thus benchmarking outcomes \cite{seedsel21}.
However, while this knowledge of that a particular covariate can possibly impact a fuzzer is a step in the right direction, in many cases these results are not actionable.
There is currently no guidance for how to  account for these covariates in future evaluations or assess their possible interactions with other variables in the evaluation setup.




\changed{In this paper, we propose two methodologies, control and randomization, which provide an \emph{actionable} framework to can account for the effects of covariates in fuzzer evaluations.
Using this framework, we suggest to report the benchmarking outcome together with the \emph{conditions under which the outcome would change}.
The first component of our methodology, a controlled experiment where one benchmark property is manipulated while all others are kept constant, can establish the degree to which changing that property \emph{causes} a change in the response variable. However, due to the large number of possible properties and the difficulty of realistically varying a property and exactly one property, the first methodology may not always be practical or possible.
Hence, we propose a second component to our methodology based on randomization and non-parametric multiple regression to identify the strength of the relationship between arbitrary benchmark properties (i.e., covariates) and the benchmarking outcome.
In doing so, one can effectively \emph{subtract} the influence of covariates without controlling their values.}

Our methodologies allow users to evaluate the degree to which the benchmarking outcome is influenced by differences of the fuzzers (which is the focus of the evaluation) \emph{as a function} of the properties of the benchmark.
For fuzzer developers, we see our approach as a way to gain insights into the variables which influence the effectiveness of their tools.
\changed{For fuzzer evaluators, we believe our framework provides the tools to make knowledge of important covariates \emph{actionable} by \emph{subtracting} their influence in realistic (uncontrolled) benchmarking scenarios.}

\paragraph{Control} Our first methodology is to control for the effect of a specific benchmark property.
We suggest to keep all other properties fixed while varying only the property of interest.
In doing so, we can directly identify the contribution of that property to the benchmarking outcome.
To illustrate this approach, we conduct two instantiating controlled experiments to evaluate the impact of program execution time and seed-corpus origin on the code coverage of various fuzzers.
In these studies, we find that both variables can have a statistically significant impact on the relative effectiveness of fuzzers.
Furthermore, we see that these differences are large enough to result in changes in the final ranking of fuzzers in practice.
For example, we observe that SymSan~\cite{chen2022symsan, chen2022jigsaw} is significantly worse at covering new control flow edges than AFL~\cite{afl} or Eclipser~\cite{choi2019grey} on a particular program.
But, it is significantly \emph{better} than either tool when that program's execution speed is \emph{only 100ms slower}, with all other variables held constant. 
We also see that initializing the starting corpus from prior fuzzing runs of different fuzzers~\cite{afl, libfuzzer} can result in different benchmarking outcomes;
changing the probability AFL will achieve higher coverage than LibFuzzer from $p=0.85$ to $p=0.15$, on the \texttt{lcms} benchmark program, for example.
\changed{However, it may not always be practical, possible, or realistic to manipulate exactly one benchmark property at a time. For instance, changing only the coverage of the initial seed corpus while keeping the number of seeds may not be possible; inducing an artificial, random slowdown during the execution of a program may not be realistic. Hence, we propose a second methodology.}

\paragraph{Randomization} Our second methodology is to randomize the benchmark configuration and apply non-parametric \emph{multiple linear regression}.
Here we suggest to vary all properties simultaneously and measure the impact of individual properties using the coefficients from a multiple linear regression.
By including the choice of fuzzer as an interaction term, we can quantify the relative impact a property has on a particular fuzzer and compare it to the effect of that property on other fuzzers.
Suppose on the current benchmark LibFuzzer beats AFL in code coverage achieved after 24 hours.
Our model can determine the degree to which a benchmark property must change (e.g., by choosing larger programs), such that AFL improves to beat LibFuzzer.
Using this methodology, benchmark maintainers and fuzzer developers will understand \textbf{when} some fuzzers perform better or worse in certain contexts and can account for confounding variables in their evaluations.
We illustrate this methodology using a modification of the FuzzBench evaluation platform, choosing eight benchmark properties that we reasonably believed to impact the fuzzer ranking for our instantiating experiment.

\vspace{0.2cm}
\noindent
\textbf{Contributions}. Concretely, this paper makes the following contributions:
\begin{enumerate}
  \item We argue that the outcome of a benchmarking procedure depends on the specific properties of the benchmark. \changed{Hence, we recommend to augment the outcome with a counterfactual analysis, specifying the conditions under which the outcome would change, to increase the soundness and utility of the evaluation.}
  \item We propose a framework for quantifying the impact of benchmark properties on the outcome \changed{using control if practical or randomization otherwise}. Specifically, we present a novel application of non-parametric regression analysis to keep the number of required experiments comparable to the traditional evaluation methodology. 
  \item We instantiate this methodology in three experiments, demonstrating its application in a practical setting. As a byproduct of this illustration, we identify several novel variables which can have significant impacts on benchmarking outcomes such as the corpus origin, program execution speed, and initial corpus coverage.
  \item We modify a leading benchmarking framework~\cite{metzman2021fuzzbench} to facilitate randomized experiments of arbitrary corpus properties, which we release publicly on acceptance.
\end{enumerate}

\noindent
In our instantiating experiments, we showcase our approach by answering four illustrative research questions.
More specifically, using two \emph{controlled} experiments, we investigate the following questions:

\begin{description}
\item[\textbf{IRQ1}] What is the impact of program execution speed on fuzzer effectiveness?
\item[\textbf{IRQ2}] What is the impact of seed corpus origin on fuzzer effectiveness?
\end{description}

Next, we utilize \emph{randomization} through the application of our non-parametric regression analysis, answering the following questions:

\begin{description}
\item[\textbf{IRQ3}] How is fuzzer ranking affected by varying a combination of properties?
\item[\textbf{IRQ4}] Does seed initial corpus coverage affect the relative ranking of fuzzers?
\end{description}





\section{Control: Measuring the Dependence of the Benchmark Outcome on One Property}

\begin{figure*}
    \centering
    \includegraphics[width=\textwidth]{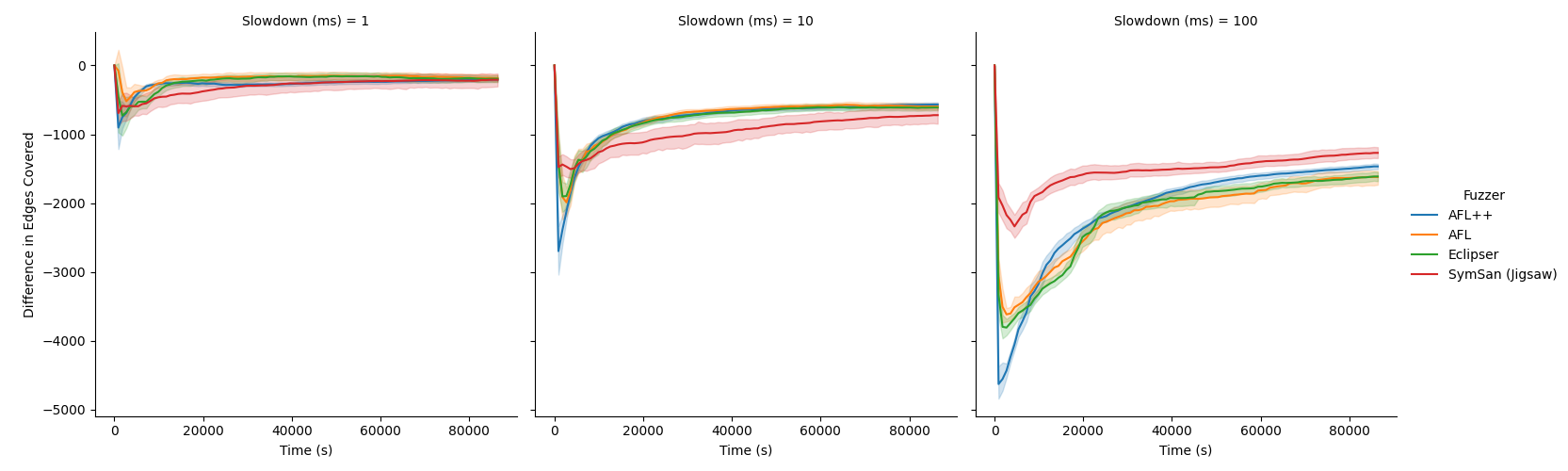}
    \caption{Decrease in coverage relative to baseline at varying slowdowns of the target program}
    \label{fig:slowdowns}
\end{figure*}

Controlled experiments are the gold-standard for empirical research.
By holding all other variables constant, researchers can estimate the impact of an explanatory variable by recording its correlation with a response variable of interest.
In the context of fuzzing, a typical benchmarking setup is an example of a controlled experiment.
All other variables --target programs, seed corpus, hardware etc.-- are controlled while the explanatory variable, the fuzzer itself, is changed.
The effect of each fuzzer is then evaluated with respect to different response variables, such as an ability to find bugs or code coverage.
Because other variables (\emph{covariates}) are controlled, the effect observed at the response variable is explained only by the effect induced by changing the explanatory variable.

However, if covariates do have an effect on the response variable, it can be important to understand what those effects are, especially if they interact with the explanatory variable.
Control only allows researchers to isolate the effect of an explanatory variable for a \emph{single configuration} of all other covariates. 
For example, if we fix the seed corpus to be empty in a controlled experiment, we might find that choosing AFL over LibFuzzer (explanatory variable) correlates with attaining higher code coverage (response variable)\cite{seedsel21}.
Because we controlled other variables, and we know that changing the fuzzer itself should change coverage achieved. 
We can even claim this to be a \emph{causal} relationship; choosing AFL \emph{causes} an increase in code coverage when the corpus is empty on the programs evaluated.
In the following section, we will use controlled experiments to understand two previously unstudied covariates in fuzzing.

\subsection{IRQ 1: Program Execution Time}

\paragraph{Motivation}
Much emphasis has been made in research on making fuzzers fast \cite{nagy2019full, gopinath2019building, schumilo2021nyx}; however, as we shall see, program execution time typically dominates greybox fuzzing campaigns.
Different programs will have different runtime characteristics, including execution time.
Thus it is important to understand how changes in the dominant portion of the fuzzing loop might affect existing fuzzers.

\subsubsection{Experimental Setup}

\paragraph{Fuzzers} We select AFL \cite{afl}, AFL++ \cite{aflpp}, Eclipser \cite{choi2019grey} and SymSan \cite{chen2022symsan, chen2022jigsaw} as our four fuzzers.
We select these fuzzers because they represent both state-of-the-art greybox fuzzers and varying levels of whitebox (symbolic) approaches to fuzzing.
We hypothesize that different fuzzers, especially those using fundamentally different approaches, will be affected differently by faster or slower executing programs.
\paragraph{Benchmark} We fix the benchmark to a single program and fuzzing harness --\texttt{libxml\_xml}-- randomly selected from those included in the Fuzzbench \cite{metzman2021fuzzbench} evaluation framework for a controlled experiment.
Similarly, we use the default seed corpus for all 20 of our 24 hour trials for each fuzzer.
\paragraph{Hardware} We use a Intel(R) Xeon(R) CPU E5-2660 v4 @ 2.00 GHz for all experiments.
\paragraph{Explanatory and Response Variables} To vary the program execution time, we modify the fuzz driver for each fuzzer to inject a delay of 100ms, 10ms or 1ms in each iteration of the fuzzing loop.
As our response variable, we investigate the delays' impact on code coverage.

\subsubsection{Results}

\autoref{fig:slowdowns} shows the net loss in edge coverage for each fuzzer at varying slowdowns of the target program.
We can see that at a slowdown of only 1ms per execution (left), the fuzzers are all similarly affected, losing roughly the same amount of coverage each when compared to no slowdown.
At a 10ms slowdown per execution (middle), again, all fuzzers lose roughly the same amount of edge-coverage. 
At this execution speed, we can also see that the greatest loss in edge coverage from slowing down the program for all fuzzers is at the beginning of a campaign, where there is a very steep dropoff in coverage.
This gap then narrows approximately logarithmically over time for all fuzzers.
When the program execution is slowed down by 100ms (right) we finally see a noticeable separation between the edge-coverage loss of each fuzzer.
During the early stages of fuzzing when the edge-coverage loss is the largest, SymSan is substantially less affected by the slowdown relative to other fuzzers.
By the 24 hour mark, SymSan remains the least affected by the slowdown, but only by a small margin over AFL++.
While we see differences in the relative effect of program execution time on each fuzzer, from \autoref{fig:slowdowns} it is not clear whether these differences will significantly affect evaluation outcomes.

\begin{figure}\centering
\scalebox{0.95}
{
\footnotesize
\begin{minipage}{0.48\columnwidth}
\begin{tabular}{@{}|@{ }l@{ }r@{ }r@{ }r@{ }r@{ }|@{}}
\hline
\input{nsections/figures/slowdown_baseline}
\\\hline
\end{tabular}\\[0.3cm]
\begin{tabular}{@{}rl}
\multicolumn{2}{@{}l}{\emph{Ranking}}\\
1. & AFL++\\
2. / 3. & AFL \\
 & Eclipser (Eclps) \\
4. & SymSan (SymS) \\
\end{tabular}\\[0.2cm]
\small
(a)~\textbf{Outcome of normal fuzzing run.}
\end{minipage}
\quad
\begin{minipage}{0.48\columnwidth}
\begin{tabular}{@{}|@{ }l@{ }r@{ }r@{ }r@{ }r@{ }|@{}}
\hline
\input{nsections/figures/slowdown_max}
\\\hline
\end{tabular}\\[0.3cm]
\begin{tabular}{@{}rl}
\multicolumn{2}{@{}l}{\emph{Ranking}}\\
1. & AFL++\\
2. & SymSan (SymS) \\
3. / 4. & AFL \\
 & Eclipser (Eclps)
\end{tabular}\\[0.2cm]
\small
(b)~\textbf{Outcome if program was 100ms slower.}
\end{minipage}}\\[0.4cm]

\caption{Pair-wise Vargha-Delaney $\hat A_{12}$ effect size between edge-coverage of fuzzers. Bold values indicate significance at $p<0.05$.\vspace{-0.2cm}}
\label{fig:slowdownvda12}
\end{figure}

\autoref{fig:slowdownvda12} (a) and (b) show the pair-wise Vargha-Delaney $\hat A_{12}$\cite{arcuri2014hitchhiker} effect size between fuzzers with no slowdown (a) and with a slowdown of 100ms (b).
Entries marked in \textbf{bold} are statistically significant by 
the Mann-Whitney $U$ test ($p < 0.05$), adjusted for multiplicity (c.f. \autoref{sec:threats}).
For example, looking at \autoref{fig:slowdownvda12} (a), the probability that a random AFL run outperforms a random SymSan run is $\hat A_{12}=83\%$.
Below the effect sizes, we list the rankings of each fuzzer by final edge-coverage.
It is clear from the rankings that the relative advantage gained by SymSan over the other fuzzers on slow programs is enough to completely change benchmarking outcomes.
In terms of effect size, SymSan goes from only having a 12\% chance of beating Eclipser in a fuzzing run to a 90\% chance on a slower program!
Similarly, SymSan flips the odds against AFL for slower programs by roughly 68 percentage points.
However, despite SymSan being relatively less affected by the slowdown than AFL++, this difference was not enough to make up for the large margin in edge-coverage achieved between AFL++ and SymSan with no slowdown.

We suspect that SymSan is the least affected by the slowdown because it is a partially symbolic approach; SymSan spends considerable time during the campaign solving path constraints with an SMT solver \cite{chen2022jigsaw} and not executing the program itself.
While Eclipser also focuses on constraint solving, it only does so in a lightweight, approximate manner for linear and monotonic constraints without invoking a solver \cite{choi2019grey}.
This result may encourage renewed interest in symbolic techniques, which have generally struggled in recent years to outperform their simpler, but highly optimized greybox fuzzing counterparts on public benchmarks~\cite{liu2023sbft}.
By expanding common fuzzing benchmarks to include slower programs, we might see a resurgence of these symbolic fuzzers.
\changed{However, further research is needed to determine if these results generalize beyond our controlled experiment.}

\result{Program execution speed \changed{can have} a significant effect on fuzzer efficacy, in absolute and relative terms. The symbolic fuzzer SymSan scales better as program execution time increases when all other variables are held constant \changed{for the \texttt{libxml\_xml} benchmark.}}

\subsection{IRQ 2: Corpus Origin}
\label{sec:origin}

\paragraph{Motivation}

It is not uncommon to see evaluations where the corpus generated in an initial campaign of one fuzzer bootstraps a larger evaluation (e.g. \cite{guler2020cupid}).
If using seeds generated by such a bootstrapping campaign can bias fuzzer evaluations, this is an important, unaccounted for threat to the external validity of these research studies.
Thus we propose a controlled experiment to determine whether the initial corpora origin affects fuzzers in practice.

\subsubsection{Experimental Setup}

\paragraph{Fuzzers} We select AFL \cite{afl}, AFL++ \cite{aflpp}, LibFuzzer \cite{libfuzzer} and Entropic \cite{bohme2020boosting} 
as representative, state-of-the-art, general purpose greybox fuzzers for our controlled experiment from two different lineages (AFL and LibFuzzer).
Our hypothesis is that corpora generated by the same fuzzer, or another similar fuzzer, will disadvantage that fuzzer in an evaluation against distinct fuzzers.
\paragraph{Benchmark} We fix the benchmark to a single program and fuzzing harness (\texttt{lcms-2017-03-21}) randomly selected from those included in the Fuzzbench\cite{metzman2021fuzzbench} evaluation framework for a controlled 24 hour experiment (20 trials).
\paragraph{Hardware} We use a Intel(R) Xeon(R) Gold 6258R @ 2.70 GHz for all experiments.
\paragraph{Explanatory and Response Variables} To vary the seed corpus origin, we sampled the starting corpus for each trial from two pools of seeds, one generated by AFL and one generated by LibFuzzer.
To create these seed pools, we conducted two pre-fuzzing runs of 24 hours each using AFL and LibFuzzer respectively.

\subsubsection{Results}


\begin{figure}\centering
\scalebox{0.95}
{
\footnotesize
\begin{minipage}{0.48\columnwidth}
\begin{tabular}{@{}|@{ }l@{ }r@{ }r@{ }r@{ }r@{ }|@{}}
\hline
\input{nsections/figures/lcms-2017-03-21.e1v0-libfuzzer.txt}
\\\hline
\end{tabular}\\[0.3cm]
\begin{tabular}{@{}rl}
\multicolumn{2}{@{}l}{\emph{Ranking}}\\
1. & AFL++\\
2. & Entropic (EnLF)\\
3. & AFL\\
4. & LibFuzzer (LF)
\end{tabular}\\[0.2cm]
\small
(a)~\textbf{Outcome if started on Lib\-Fuzzer-generated seeds.}
\end{minipage}
\quad
\begin{minipage}{0.48\columnwidth}
\begin{tabular}{@{}|@{ }l@{ }r@{ }r@{ }r@{ }r@{ }|@{}}
\hline
\input{nsections/figures/lcms-2017-03-21.e1v0-afl.txt}
\\\hline
\end{tabular}\\[0.3cm]
\begin{tabular}{@{}rl}
\multicolumn{2}{@{}l}{\emph{Ranking}}\\
1. & Entropic (EnLF)\\
2. & LibFuzzer (LF)\\
3. & AFL++\\
4. & AFL
\end{tabular}\\[0.2cm]
\small
(b)~\textbf{Outcome if started on AFL-generated seeds.}
\end{minipage}}\\[0.4cm]

\footnotesize
\begin{tabular}{@{}r|rrrr@{}} 
\input{nsections/figures/lcms-2017-03-21.effect.txt}
\end{tabular}\\[0.2cm]
(c)~\textbf{Intra-fuzzer effectiveness difference  if the evaluation was started with AFL-generated seed corpora instead of Lib\-Fuzzer-genera\-ted ones.}
\caption{Vargha-Delaney $\hat A_{12}$ effect size of edge-coverage between fuzzers. Bold values indicate significance at $p<0.05$.\vspace{-0.2cm}}
\label{fig:seedorigin}
\end{figure}

\autoref{fig:seedorigin} shows the outcome of this experiment.
The grids in subfigures (a) and (b) shows the pair-wise effect size measured using Vargha-Delaney's $\hat A_{12}$, highlighted in bold for statistical significance according to the Mann-Whitney $U$ test (p < 0.05), adjusted for multiplicity (c.f. \autoref{sec:threats}).
On this program, with initial corpora generated by LibFuzzer, AFL++ performed best and LibFuzzer worst.
However, the evaluation outcome is very different if we run those same fuzzers on corpora generated by AFL.
\autoref{fig:seedorigin}.b shows the ranking \emph{if AFL-generated seeds were provided} instead. Now Entropic performs best while AFL performs worst. The probability that an arbitrary AFL run outperforms an arbitrary LibFuzzer run is $\hat A_{12}<1\%$. \emph{The choice of seed origin has a substantial impact on the benchmark outcome}. \autoref{fig:seedorigin}.c further demonstrates this impact. All fuzzers perform worse on AFL-generated seeds ($\hat A_{12}<18\%$). Yet, the impact on AFL/AFL++ is greatest ($\hat A_{12}<8\%$): AFL-based fuzzers perform worse using AFL-generated seeds.

In addition to showing that the corpus origin \changed{can have} a significant impact on evaluation outcomes, we can also begin to see some trends with this controlled experiment.
Both fuzzers from the LibFuzzer lineage (LibFuzzer and Entropic) improve in ranking when run on AFL-generated corpora.
Similarly, the fuzzers from the AFL lineage (AFL++ and AFL) improve on LibFuzzer-generated corpora.
In other words, using seed inputs generated by a fuzzer not only can negatively impact that fuzzer, but also may negatively impact \emph{other similar fuzzer implementations}.
Given that many fuzzers presented in research are modifications of existing tools~\cite{aflfast,aflpp,bohme2020boosting}, this means that any evaluation utilizing fuzzer-generated seeds could be biased by this behavior.
One explanation for this result is that similar fuzzers tend to explore similar behaviors in the program under test.
If most of those behaviors are already covered by the initial corpus, it leaves less room for the fuzzer to explore during the evaluation.

\result{Corpus origin \changed{can have} a significant influence on fuzzer effectiveness. \changed{On the \texttt{lcms} benchmark program,} we observe that tools used to generate the initial corpora for an evaluation are negatively affected in that evaluation. Even distinct fuzzer implementations based on the same lineage as the the initial corpus generator can be significantly negatively affected.}

\subsection{Challenges of Control}
\changed{While preferable, our methodology of control is not always applicable.
Our controlled experiments demonstrate that the result of the evaluation \emph{can change} depending on the specific configuration of the benchmark.
However, there are many issues that can arise when applying this methodology in practice.
In particular, there are two core reasons why a second methodology may be needed instead of or in conjunction with controlled experiments.}

\paragraph{Specificity}
\changed{In order to keep other variables like program size and execution time constant, the fully controlled experiments in \textbf{IRQ1} and \textbf{IRQ2} were only run for a single benchmark program and configuration. Thus, their results cannot be safely generalized beyond this configuration without additional experiments.
Does execution time affect fuzzers on programs other than \texttt{libxml}?
We can test our hypothesis on other benchmark programs, but this necessarily means that our setup is no longer fully controlled; now \emph{both} the execution speed and the program itself are variables that may affect the response.
If we naively aggregate the results on these different programs, the resulting variance will include the effects of the programs themselves, reducing our ability to distinguish statistically significant effects for our variable of interest.
We also gain no insights into what general program characteristics might impact our response variable.
For example, programs which are already very slow might be less impacted by a further reduction in execution speed, but a single aggregate statistic over several programs cannot capture this relationship.
The methodology we propose in the remainder of this paper can aggregate data from multiple programs and give a nuanced view of program characteristics with a unified approach.
}

\paragraph{Dependence and Realism}
\changed{Unlike the choice of seed origin, many benchmark properties are difficult or impossible to manipulate independently.
For example, it might not be possible to manipulate the size of the fuzzer seed inputs while holding the seeds' validity, the coverage of the corpus, or the execution time of the program constant.
In answering \textbf{IRQ1}, we manipulated execution time without otherwise modifying the instructions executed or control flow and while holding all other properties constant. 
This allows us to make strong claims about execution speed specifically having different effects on different tools.
However, in practice, execution speed is often (but not \emph{always}) a product of the number of instructions and branches in a program.
Additional instructions and branches translate into larger, more complex formulae passed to an SMT solver for symbolic approaches.
Thus, a program that is slower \emph{because} of additional complicated control flow could easily result in SymSan performing relatively worse than AFL -- the exact opposite of the results from our controlled experiment in \textbf{IRQ1}!
Yet, increasing the number of instructions or branches in a program while holding all other variables constant is, again, extremely challenging: What if the locations of the inserted code has an impact? What if the inserted code impacts the reachability of parts of the existing program? Which instructions or branch conditions should be inserted to make the setup ``realistic''?
Rather than attempt to construct a completely controlled experiment that may end up being highly artificial or impractical, we give a secondary methodology for these scenarios that allows us to assess the impact of variables \emph{without} controlling their values.
This also makes findings by prior work (e.g. \cite{seedsel21}) actionable, in that the influence of known or suspected covariates can be effectively \emph{subtracted} from experimental outcomes to obtain an unbiased view of fuzzer performance.}

\section{Randomization: Measuring the Dependence on Multiple Benchmark Properties}\label{sec:holistic}
When control is not practical or possible, our fuzzer evaluation methodology consists of a \emph{non-parametric regression analysis} of benchmarking data, using \emph{bootstrapped confidence intervals} to test for statistical significance of results.
By constructing a regression model, one can account for the effects of multiple variables simultaneously with respect to a performance metric.
As our technique does not assume a particular error or data distribution, it is widely applicable in tool evaluations, even on non-linear data with strong outliers.
Finally, because regression models form the basis for ANOVA and hypothesis testing, our approach can be used to make rigorous claims about the statistical significance of results, in addition for being useful for exploratory analysis.
To enable our holistic methodology, we propose to randomize, rather than control, benchmark properties for each run of the evaluation and record the properties of interest.
For example, controlling initial corpus by holding it static across repetitions, as in traditional evaluation, reduces variance. However, it excludes effects of the corpus on the benchmarking outcomes in doing so.\vspace{-0.1cm}

\paragraph*{Randomization}
Unlike in typical fuzzing evaluations, we recommend randomly varying properties of the benchmark for each repetition of the experiment.
For example, we can accomplish this for \emph{corpus properties} by sampling each starting corpus from a larger pool of seed inputs.\footnote{
In this case, B\"{o}hme and Falk \cite{bohme2020expfuzzing} observed that a linear increase in coverage requires an exponential increase in the number of inputs.
Hence, the \emph{average} corpus, if sampled uniformly, might still be saturated.
Instead, we recommend to sample from an exponential distribution centered at a desired percentage of the saturated corpus.}
We also suggest using a \emph{matched-pairs} experimental design, such that each fuzzer is run on each configuration (e.g. starting corpus) to allow for direct comparisons.
Other benchmark configuration parameters, such as the benchmark programs themselves, compilers used and architecture can also be randomized or enumerated where practical.\vspace{-0.1cm}

\paragraph*{Rank transformation} For holistic fuzzer evaluation, we recommend applying the rank transformation to our gathered data, replacing each data value with its relative rank in the overall dataset.
For example, the trial with the lowest coverage on a given program would have a rank-transformed value of 1, the second lowest 2, etc.

There are several reasons for studying the \emph{ordinal association} rather than the numeric association between benchmark property and performance measure. First, we do not need to assume that the distribution of the data itself is normal or that the relationship between our explanatory and response variables is linear. In fuzzing we often observe extreme outliers or exponential effects which provide undue influence on a measure of the strength of the association (i.e., correlation).
Secondly, we do not require value domains to substantially overlap across programs. For instance, we observe that a low initial coverage for one program can be a high initial coverage for another. Hence, we conduct our analysis on the the rank-transformed value \emph{within} a program for corpus properies or across all programs w.r.t. program properties.
Finally, rank transformations are used in non-parametric methods when values vary widely in scale \cite{kruskalwalis,iman1979use,mannwhitney,nonparametrics,conover1981rank}---like in automatic software testing. For instance, the Friedman test \cite{iman1979use} describes a similar regression analysis against the ranks.\vspace{-0.1cm}

\paragraph*{Regression Analysis} To quantify the combined effect of benchmark properties and fuzzers on benchmarking outcome, we propose a \emph{multiple linear regression} of rank-transformed data as our holistic benchmarking technique. A linear regression model represents a response variable, e.g. coverage or fuzzer ranking, as a linear combination of several explanatory variables.
Each regression coefficient can be considered as a measure of effect size for its corresponding variable: it indicates the degree to which that explanatory variable contributes to the response variable. 

In the case of fuzzer benchmarking, we can give a general formula for a generic set of fuzzers, benchmarks, and properties. Let $P=\{p_i\}_{i=1}^n$ be a set of $n$ benchmark properties and $F$ be a set of fuzzers. An example of $p\in P$ is program size. We first choose a \emph{reference configuration} $C=\langle f,\{v_i\}_{i=1}^n\rangle$ by selecting a \emph{reference fuzzer} $f\in F$ and \emph{reference values} $v_i$ for every property $p_i\in P$.
A multiple linear regression finds the intercept $\alpha$ and the regression coefficients $\beta_p$, $\gamma_f$, and $\omega_{p,f}$, such that
\vspace{-0.2cm}

\begin{align}
\label{eqn:holistic}
R &= 
  \alpha + \left[\sum_{p_i\in P} \beta_i X_i\right] + \left[\sum_{f\in F} \gamma_f Y_f\right] + \left[\sum_{p_i\in P}\sum_{f\in F} \omega_{i,f} X_i Y_f\right]
\end{align}%
where $X_i$ is the rank of the $i$-th property relative to the reference level $v_i\in C$, and where $Y_{f}\in \{0,1\}$ are indicator variables such that $Y_{f}=1$ if $f\in F$ was used instead of the reference fuzzer $f'\in C$. 
By fitting the regression model on  benchmarking data, one obtains estimated values for each of the coefficients in the model, and the thus effect size of each variable.
For instance, if $\beta_i>0$, we say that an increase in $X_i$ by one unit gives an increase in fuzzer ranking by $\beta_i$. 
Similarly, the coefficient $\omega_{pf}$ of an \emph{interaction term} describes the additional contribution of the benchmark property $X_{p}$ if the fuzzer $f$ was used instead of the reference fuzzer (LibFuzzer).

\paragraph*{Need for multiple regression}
We recommend multiple-regression because it accomplishes our two primary goals:
(1) it provides a way to conduct hypothesis testing for statistical significance and a measure of effect size and (2) can account for multiple variables in its analysis.

Regression analysis encompasses a wealth of techniques, such as a standard t-test or bootstrapping, which one can use for hypothesis testing.
Similarly, the regression coefficients themselves represent a measure of relative effect size for their corresponding variables.
Multiple regression also allows one to account for the effects of a variable as if all other variables were held constant, even if this fine-grained control is not feasible in practice.
This last point is crucial in software evaluations where various aspects of the inputs and the programs cannot be manipulated in isolation from each other.

Additionally, regression models and ANOVA are a well-understood and flexible group of techniques which are standard in many fields for statistical analysis.
While \emph{hypothesis testing} is vital for researchers to establish empirical differences between fuzzers, \emph{exploratory analysis} may be of more interest to practitioners.
Even without an \textit{a priori} hypothesis, a regression model and its corresponding coefficients can be used to approximately gauge the effects of many variables using the same experimental setup and analysis.

\paragraph*{Bootstrapped confidence intervals}
As noted by Arcuri and Briand \cite{arcuri2014hitchhiker}, the assumption that errors are distributed normally is often violated in software engineering contexts because the data itself is not normally distributed.
We address this concern by using \emph{bootstrapping} as a method for computing non-parametric confidence intervals.
By repeatedly regressing against sub-sampled data from our initial sample, we can obtain confidence intervals computationally, rather than analytically (avoiding the normality assumption).

\paragraph*{Assumptions}
Linear regression models make several assumptions about the underlying data, such as linearity, homoscedacity and independence.
We outline how to check these assumptions in \autoref{sec:assumptions}, and we guard against violating them with two other core aspects of our methodology: the \emph{rank transformation} and \emph{bootstrapping}.

\subsection{Randomization Instantiation}


To showcase our randomized holistic evaluation methodology, we \emph{instantiate} it on a set of reasonable benchmark and initial corpus properties known or suspected to impact fuzzers.
In doing so, we answer the following illustrative research questions using the data from a leading fuzzer evaluation platform, 
\begin{description}[leftmargin=0.8cm]
\item[\textbf{IRQ3}] How is fuzzer ranking affected by varying a \emph{combination} of 
properties? \textbf{(Exploratory Analysis)}
\item[\textbf{IRQ4}] Does initial corpus coverage \emph{significantly} affect the \emph{relative} ranking of fuzzers? \textbf{(Hypothesis Testing)}
\end{description}

\begin{figure}
\begin{minipage}{0.5\columnwidth}
\centering\scriptsize
    \begin{tabular}{ @{}l@{ \ }l@{}}\hline
    \textbf{Program} &\textbf{Description}\\\hline
    freetype2 & Font Renderer \\
    harfbuzz & Text Shaping \\
    libjpeg-turbo & JPEG Image Codec \\
    libpcap (fuzz\_both) & Packet Capture \\
    libxslt (xpath) & XML Transformation\\
    libpng & PNG Image Codec\\\hline
\end{tabular}
\end{minipage}\hspace{-0.1cm}%
\begin{minipage}{0.5\columnwidth}\centering\scriptsize
\begin{tabular}{@{}l@{ \ }l@{}}\hline
    \textbf{Program} &\textbf{Description}\\\hline
    sqlite3 (ossfuzz) & Embedded Database \\
    vorbis & Audio Encoding\\
    woff2 & Web Fonts \\
    zlib (uncompress) & Decompression\\
    mbedtls (dtlsclient) & Cryptographic \\
        & Primitives \\\hline
\end{tabular}
\end{minipage}\vspace{-0.25cm}
\caption{Benchmark Programs\vspace{-0.35cm}}
\label{fig:programs}
\end{figure}

\subsubsection{Experimental Setup}
\label{sec:setuprand}

\paragraph{Hypothesis (IRQ4)}
Before running an experiment to test for a statistically significance, it is important to first formulate testable hypotheses.
In this case, we use a reframing of IRQ4 as our instantiating hypotheses \textbf{H0.1-3}, but any other variable(s) in our model can be used.

\textbf{H0:} Changes in Initial Coverage have the same impact wrt. final coverage ranking on LibFuzzer as it does on... 
\begin{enumerate}
    \item  AFL,
    \item AFL++, and
    \item Entropic.
\end{enumerate}

We recommend hypothesis testing on a single or small number of predetermined variable(s) to avoid Type I error (Section \ref{sec:threats}).
For IRQ3, as we are only using the model for exploratory analysis, no formal hypotheses are needed.

\paragraph{Benchmark} (\autoref{fig:programs}).
We chose the FuzzBench \cite{metzman2021fuzzbench} platform to apply our holistic evaluation methodology because it is a leading tool with widespread usage among researchers and practitioners,\footnote{More than 140,000 CPU-days of public experiments run in 2022.} and there is large industry support from its corporate sponsor.
Fuzzbench uses \emph{branch coverage} as its measure of fuzzer performance. However, 
in principle our methodology applies to any other benchmark framework or performance measure \cite{hazimeh2020magma,profuzzbench,metzman2021fuzzbench}. 

\paragraph{Fuzzers} We chose AFL \cite{afl}, AFL++ \cite{aflpp}, LibFuzzer \cite{libfuzzer}, and Entropic \cite{bohme2020boosting} (extends LibFuzzer), representing the state-of-the-art in general-purpose grey-box fuzzing.
\paragraph{Programs} To maximize the relevance of our findings, we use all programs directly integrated into FuzzBench\footnote{Rather than adding additional programs or harnesses through the OSS-Fuzz Integration. Programs already added to Fuzzbench are widely used and thus results on these subjects are more directly relevant to evaluators} (i)~that were already available in Commit~\texttt{8858be7}, (ii)~that could be compiled and run within the local setup, and (iii)~that had an OSS-Fuzz corpus available in Fuzzbench. Our benchmark consists of 11 programs listed in \autoref{fig:programs}.
For every program, all experiments were conducted as FuzzBench local experiments on an AWS \texttt{c6i.metal} instance.

\paragraph{Benchmark properties}
For our experiments, we select a set of reasonable properties, to \emph{instantiate} our holistic benchmarking framework.
We hope that this illustrative example analysis can both provide direct insights on the properties we evaluate and exemplify how future experiments can be conducted with new properties.
We do not claim that these properties are exhaustive; we only investigate a subset of interesting properties which have been identified by prior work. As \emph{corpus properties}, we measure the {number of seeds}, the {initial LLVM branch coverage} before the fuzzing campaign, the {mean execution time} in nanoseconds, and the {mean size of the seeds} in bytes.
As \emph{program properties}, we measure the {size of the program}, the {proportion of equality and inequalities} in comparison instructions, and the {proportion of shared library calls}.

We chose seed size and execution time as covariates as the documentation for AFL claims that each have a strong impact on performance \cite{AFLsize}\cite{AFLtime}.
We also record the number of shared library calls because instrumentation is the defining characteristic of grey-box fuzzing techniques, shared libraries are not instrumented by default and are even recommended to leave uninstrumented \cite{AFLPPltoshared}.
We examine the impact of initial corpus size based off of observed effects in prior published work. 
For example, Klees et al. \cite{klees2018evaluating} show a difference between the empty and non-empty seed sets, but not for more granular changes in the size of the seed corpus.
Additionally, some fuzzers like AFLFast have known issues with very large working corpora \cite{zhu2019feature}.
We measure initial coverage because it is well established that the performance characteristics change at different stages in a fuzzing campaign \cite{bohme2020expfuzzing}.
Similarly, we include program size as it impacts both the search space itself and how the fuzzer represents that space, which can lead to other performance issues such as those highlighted by CollAFL \cite{gan2018collafl}.

\paragraph{Instrumentation}
To measure shared library calls and constraints in each program, we used \texttt{objdump} to extract the corresponding static instructions and their characteristics.
We then used a binary instrumentation tool \cite{duck2020binary} to count these calls and constraints at runtime on the initial corpus.
For program size, we report the size of the text segment in the program binary given by \texttt{objdump}.

\paragraph{Sampling configuration}
For every \{fuzzer, program\}-pair, we conducted 24 trials of 24 hours.
For each trial, we create the initial seed corpus by randomly sampling from a saturated seed corpus that was generated over many years by the OSS-Fuzz continuous fuzzing platform \cite{ossfuzz}.\footnote{We use the FuzzBench option \texttt{--oss-fuzz-corpus} and minimize with \texttt{afl++-cmin}.}
This highly saturated corpus approximates the ``universe'' of seed inputs.
For the $x^{th}$-trial ($x:1 \le x\le 24$) of each program, all fuzzers start from the same initial corpus.
Notably, we opt for a single trial per benchmark configuration.
Running multiple trials per benchmark configuration would allow us to estimate the variance \emph{not associated with the benchmark properties} themselves and compare to the variance when those properties are changed.
This additional capability, however, comes at the cost of substantial additional computational resources (e.g. a factor of 20 for 20 trials per configuration).
Using a single trial per configuration, we can still determine statistically significant results in terms of our response with respect to the overall variance.
Given that fuzzer evaluations are already extremely costly, we believe that a single trial per configuration is likely to be preferred by most evaluators and thus choose this as our experimental setup.

%



\begin{figure*}[t]
    \centering\footnotesize
    \begin{minipage}{0.195\textwidth}\centering
    \includegraphics[width=0.9\textwidth]{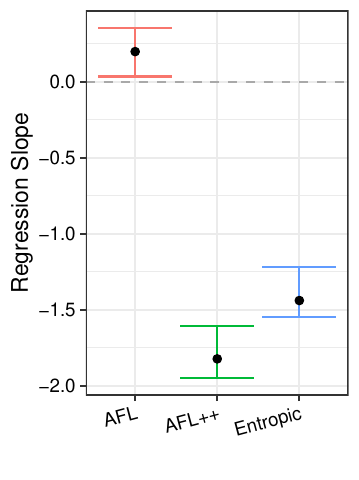}\\
    \scriptsize
    $R^2=0.65 (0.65 Adj.)$, \emph{p-val.}$<0.001$\\
    Median Residuals: $0.045$
    \end{minipage}
    \begin{minipage}{0.785\textwidth}\centering 
    \includegraphics[width=\textwidth]{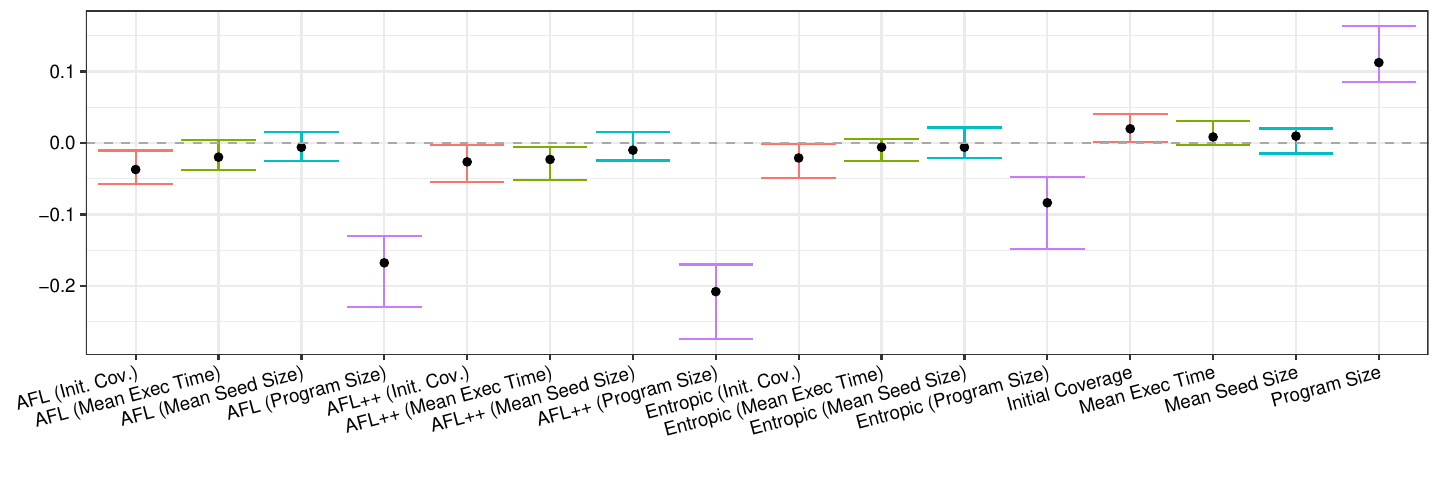}
    \end{minipage}\vspace{-0.1cm}
    \caption{Multiple Linear Regression with LibFuzzer as reference level (Fuzzer Ranking $\sim$ Fuzzer $\times$ Properties) [Eqn. \ref{eqn:holistic}\vspace{-0.1cm}].
    }
    \label{fig:mlr}   
\end{figure*}

\subsection{Results}
\label{sec:t3}

\label{sec:t3analysis}
To analyze the results of our experiment,   we instantiate the regression model introduced in \autoref{sec:holistic} with our selected corpus properties and response variable (Eqn.~\ref{eqn:holistic}).
We define the following \emph{reference configuration} $C$. As reference levels $\{v_i\}_{i=1}^n\in C$ we choose: for all corpus properties the lowest rank and for program size to the median rank. As reference fuzzer $f\in C$, we choose LibFuzzer.

%
 \autoref{fig:mlr} shows the coefficients as point estimates and the non-parametric bootstrapped 95\%-confidence intervals (CIs), adjusted for family-wise error rate as described in \autoref{sec:threats}. If the CI does not include the origin (0), we can conclude that there is a statistically significant effect from the corresponding benchmark property relative to the reference levels at that level of confidence. The left side of  \autoref{fig:mlr}  shows the change in fuzzer rank if a different fuzzer is chosen and all other covariates (i.e., properties) are held constant at the reference level. The last four whiskers in \autoref{fig:mlr} (right side) show the change in fuzzer rank if one of the four considered properties increase by one rank and the fuzzer is held constant (LibFuzzer). The remaining whiskers show the \emph{additional} change in fuzzer rank if both fuzzer and benchmark property are changed simultaneously --- interaction terms in equation (1) presented earlier.

\subsection*{IRQ3: Exploratory Analysis of Properties in Combination}
Looking at \autoref{fig:mlr}, we see several substantial effects from our explanatory variables.
On the left, we see that AFL appears to be outperformed by LibFuzzer, our reference fuzzer, at the reference levels for each of our covariates -- its confidence interval is entirely above zero.
Examining results on the right for corpus properties, we see that all three other fuzzers perform better with higher initial coverage corpora, as the confidence intervals are each below zero.
Additionally, increasing mean execution time has a negative association with the performance of LibFuzzer relative to AFL++ (and possibly AFL as well). 
Finally, each other fuzzer appears to perform better than LibFuzzer as program size increases. 

In light of these findings, we can gain some informal intuition about which fuzzers to use under which circumstances or formulate new, targeted hypotheses to focus on in future experiments.
We might use AFL++ and Entropic and AFL over LibFuzzer for programs of size comparable to the medium and larger programs in Fuzzbench (median of 4917 instructions, max of 14452 instructions).
However, on smaller programs with lower initial coverage and smaller or faster seeds, this gap in performance (coverage achieved) narrows significantly, and Entropic performs best, followed by AFL++ and LibFuzzer.
However, for such conclusions to be empirically confirmed, we would need to run a new experiment to test each hypothesis, as 
will be exemplified in \textbf{IRQ4}.

Our holistic fuzzer evaluation methodology also allows us to put the negative impact of a benchmark property \emph{in relation} to the impact of the choice of fuzzer at a reference level.
Keeping all other covariates constant at their reference levels, a switch from LibFuzzer to AFL is roughly equivalent to increasing the program size from median size programs to the largest programs in our benchmark set.
Comparing the coefficients for the fuzzer/corpus-property interactions, AFL++'s seem to be of similar magnitude to other fuzzers, suggesting that it is not excessively overfit to a particular starting corpus relative to other fuzzers.

\begin{figure*}\centering\footnotesize
\begin{tabular}{@{}l@{ }ll@{ }ll@{ }ll@{ }l|l@{ }l@{}}
\multicolumn{2}{@{}l}{\textbf{Benchmark Config 1}\quad\quad\quad} &
\multicolumn{2}{l}{\textbf{Benchmark Config 2}\quad\quad\quad} &
\multicolumn{2}{l}{\textbf{Benchmark Config 3}\quad\quad\quad} &
\multicolumn{2}{l|}{\textbf{Benchmark Config 4}\quad\quad\quad} &
\multicolumn{2}{l}{\textbf{Official FB Config}}\\
\multicolumn{2}{@{}l}{{\scriptsize $\downarrow$}\ \emph{Low Initial Coverage}} &
\multicolumn{2}{l}{{\scriptsize $\downarrow$}\ \emph{Low Initial Coverage}} &
\multicolumn{2}{l}{{\scriptsize $-$}\ \emph{Median Initial Coverage}} &
\multicolumn{2}{l|}{{\scriptsize $\uparrow$}\ \emph{High Initial Coverage}} &
\multicolumn{2}{l}{{\scriptsize $\cdot$\ }\emph{Static seed set}}\\
\multicolumn{2}{@{}l}{{\scriptsize $\downarrow$}\ \emph{Small Programs}} &
\multicolumn{2}{l}{{\scriptsize $\uparrow$}\ \emph{Large Programs}} &
\multicolumn{2}{l}{{\scriptsize $-$}\ \emph{Median Sized Programs}} &
\multicolumn{2}{l|}{{\scriptsize $\uparrow$}\ \emph{Large Programs}} &
\multicolumn{2}{l}{\ \ \emph{used by Fuzzbench}}\\
\multicolumn{2}{@{}l}{{\scriptsize $\downarrow$}\ \emph{Small and Fast Seeds}} &
\multicolumn{2}{l}{{\scriptsize $\downarrow$}\ \emph{Small and Fast Seeds}} &
\multicolumn{2}{l}{{\scriptsize $-$}\ \emph{Median Size and Speed Seeds}} &
\multicolumn{2}{l|}{{\scriptsize $\uparrow$}\ \emph{Large and Slow Seeds}} &
\multicolumn{2}{l}{\ \ \emph{in all prior work}}\\[0.1cm]\hline
& & & & & & & & & \\[-0.2cm]
1. & Entropic &
1. & AFL++ &
1. & AFL++ & 
1. & AFL++ & 
1. / 2. & AFL++ / Entropic\\
 & LibFuzzer &
 & Entropic &
2. & Entropic &
2. & Entropic &
2. / 3. & Entropic / AFL\\
3. & AFL++ &
3. & AFL &
3. & LibFuzzer &
 & AFL &
\\
4. & AFL &
 & LibFuzzer &
4. & AFL &
4. & LibFuzzer &
4. & LibFuzzer\\
\end{tabular}

\caption{(left) Benchmarking outcomes at various levels of program and corpus properties (significant at bootstrapped 95\% CI), (right) Benchmarking outcome from the Fuzzbench default corpora (significant at p < 0.05, Mann-Whitney U-test)}
\label{fig:rankings}
\end{figure*}

\autoref{fig:rankings} shows the rankings given by our models at varying values of our measured corpus and program properties.\footnote{Differences in rankings is determined by the bootstrapped 95\% confidence intervals, adjusted for multiplicity (c.f. \autoref{sec:threats})}
These rankings can easily be obtained from our model by adjusting the reference levels and fuzzers of \autoref{fig:mlr}.
The rankings are notably different depending on the values of our measured properties.
LibFuzzer ranges from being among the best fuzzers (\autoref{fig:rankings}, far left) to the worst (\autoref{fig:rankings}, right).
Similarly, AFL++, which has been highly optimized against the static Fuzzbench corpus falls as low as third.
Furthermore, these rankings are only slices at discrete points in the space of possible evaluation configurations.
One of the strengths of our holistic model is that it can produce such rankings at any point in the space, along with confidence intervals to gauge the significance of these results.

In contrast, the traditional evaluation setup can only give rankings for precisely one point in the space of configurations.
We show the Fuzzbench \cite{metzman2021fuzzbench} rankings for the four fuzzers we tested on the right of \autoref{fig:rankings}.
We compute the fuzzer rankings from the average run of each fuzzer across all our benchmark programs in existing experimental data, using the Mann-Whitney U-Test for significance (p < 0.05) \cite{fbpaperdata}.
The Fuzzbench rankings only represent the relative performance of these fuzzers for one initial corpus per program -- the default seeds provided by Fuzzbench.
The choices of initial corpora utilized by Fuzzbench are not explained, and seem to be arbitrary.
Several target programs such as \verb|jsoncpp| and \verb|libjpeg| start with only a single input, yet others like \verb|sqlite3| start with more than one thousand inputs.
Given that we observe corpus properties have a significant effect on evaluation outcomes, this heavy usage of a single configuration is likely to introduce bias into the results.
Indeed, we can see from the far right of \autoref{fig:rankings}, that the rankings output by Fuzzbench are not the same as the rankings for representative median values of the properties that we observed in our study (\autoref{fig:rankings}, middle).
One could say that the Fuzzbench results (and thus fuzzers tuned on these results) may be \emph{overfitted} to the Fuzzbench default seed set.
The potential for this bias reinforces the need for evaluation platforms to utilize a wide variety of sampled starting corpora and benchmark programs as in this paper, rather than arbitrarily choosing a single evaluation configuration.

\result{Our holistic model shows potential effects from the \textbf{program size} and \textbf{initial coverage} of the seed corpus, as well as some effects from the execution time of the seeds in the starting corpus on some fuzzers.
These effects indicate how the final ranking of a fuzzer would change holding all other variables constant.
Large variances in these properties appear to be enough to change fuzzer rankings in practice.
}

\subsection*{IRQ4: Hypothesis Testing}
\label{sec:hyp}

Instead of an exploratory analysis, we can leverage our holistic model to do hypothesis testing.
In this case, we use \textbf{H0.1-3} as our instantiating hypotheses, however the impact of any property could be tested in this way.
Before testing multiple individual hypotheses, we first run ANOVA for our model to see if \emph{any} variable effects are significant.
Here we see a P-value of far less than our chosen significance level of $0.05$ (\autoref{fig:mlr}, left) and so we conclude that at least one of the variables in our model contributes significantly to our response variable.

To run a post-hoc test for a hypothesis using our holistic model, we can simply see if the confidence interval for the coefficient of the corresponding variable overlaps with zero.
In this case we are looking at the relative effect of the initial coverage of the seed corpus for each fuzzer against LibFuzzer, which is captured by the interaction terms of (fuzzer*seed size).
Looking at \autoref{fig:mlr}, we can see that \emph{none} of the confidence intervals reach zero.
Thus we can reject the null hypothesis for \textbf{H0.1-3} and claim that initial coverage has a significantly different effect on other fuzzers relative to LibFuzzer.

To test other hypotheses between other fuzzers (e.g. AFL vs. Entropic), we could simply replot the data such that e.g. AFL, rather than LibFuzzer is the reference fuzzer.

\result{The bootstrapped confidence intervals from our holistic regression model can be used to test individual hypotheses for significance.
In our instantiating experiment, we would reject the null hypotheses and conclude that \textbf{initial coverage} has a statistically significant relative impact on AFL, AFL++, and Entropic relative to LibFuzzer.
}

\subsection{Assumptions and Model Validation}
\label{sec:assumptions}

To ensure the correctness of our approach, we examine the fit and prediction accuracy of our regression model.
We also check the general assumptions of regression analysis.

\subsubsection{Model Validation}
\label{sec:valid}

\begin{figure}\centering
\footnotesize
\begin{tabular}{ l r r r r }
\toprule
        Model & Acc. (\%) &  $R^2$ & Adj. $R^2$ & DoF \\
\midrule
        Atomistic (static rankings) & 49.6 & 0.484 & 0.485 & (3, 790) \\
        Holistic [Eqn. \ref{eqn:holistic}] & 59.5 & 0.669 & 0.661 & (19, 774) \\\midrule[0.01pt]
        Extended Holistic [Eqn. \ref{eqn:bench}] & 67.6 & 0.780 & 0.763 & (59, 734) \\
        
\bottomrule
\end{tabular}\\
\caption{Prediction accuracy and model statistics \vspace{-0.2cm}}
\label{tab:r2}
\end{figure}

To evaluate the fit of our model we examined the prediction accuracy and mean-squared error relative to an atomistic model.
\changed{The atomistic model predicts that a fuzzer's ranking on the training set will correspond to its ranking on the holdout set. 
This atomistic model represents the current state-of-the-practice in fuzzer evaluation -- i.e. the average rank shown at the top of a Fuzzbench report.\footnote{\url{https://www.fuzzbench.com/reports/sample/index.html}}}
We use a 75\%-25\% train-test split of the data from the previous section for evaluating prediction accuracy of our response variable, with 5-fold cross-validation on the training set.

\autoref{tab:r2} shows the holdout-set prediction accuracy, goodness of fit ($R^2)$, and degrees of freedom (DoF) for the traditional atomistic approach and our proposed holistic model.
The traditional approach in the first row mirrors how users choose fuzzers today.
This model chooses a ranking for unseen data based only on the overall rankings on prior data (the best fuzzer overall is always predicted to have ranking 1, the second best 2, etc.).
For the holistic model in the second row, our multiple linear regression model (fitted on the seen benchmark configurations) predicts the benchmark outcome on an unseen configuration.
We use the extended holistic model in the third row of \autoref{tab:r2} to assess variance from program-specific effects, and will explain its precise nature in the coming section.

Given an unseen benchmark configuration, our holistic model can predict the correct ranking of a fuzzer roughly $60\%$ of the time.
In contrast, an atomistic model can only predict a fuzzer's ranking correctly less than half of the time on the unseen configurations.
The holistic perspective thus gives users a substantially better chance of selecting a fuzzer that best fits their use case.


\result{Our holistic methodology is nearly $20\%$ more accurate in predicting the correct ranking of a fuzzer than a traditional evaluation ($10$ percentage points).}

Looking at the goodness-of-fit ($R^{2}$), we see the holistic model is substantially better than the traditional approach (+0.18 units).
We can also remove the benchmark properties from our holistic model to assess their cumulative impact on evaluation outcomes in terms of $R^2$ value. 
Doing so creates a ``fuzzer-only'' regression model, which has an $R^2$ of 0.604 units.
In other words, the properties explored in our investigation explain an additional $6.5$ percentage points of the total variance in fuzzer rankings.
However, 33\% of the total variance in predicting the fuzzer ranking $R$ remains unexplained by our model.

From the remaining variance, we hypothesize that there are significant program-specific effects not measured in our study, and thus not captured in our holistic model.
By adjusting our model to take into account the benchmark program as a categorical variable, we can incorporate some of these effects in the aggregate. Let $B$ be the set of programs. For each $b\in B$ and $f\in F$, let $\mu_b$ and $\rho_{b,f}$ be additional coefficients. Given reference level $C'=\langle f,b,\{p_i\}_{i=1}^n\rangle$, the fuzzer rank $R$ is

\vspace{-0.3cm}
{\small%
\begin{align}
\label{eqn:bench}
R =  
  \alpha &+ \left[\sum_{p\in P} \beta_p X_p\right] + \left[\sum_{f\in F} \gamma_f Y_f\right] + \left[\sum_{p\in 
 P}\sum_{f\in F} \omega_{p,f} X_p Y_f\right]\\
%
&+ \left[\sum_{b\in B} \mu_b Z_b\right] + \left[\sum_{b\in B}\sum_{f\in F} \rho_{b,f} Z_b Y_f\right]\nonumber
\end{align}%
}%
%
Here, $Z_b$ is an indicator variable, s.t. $Z_b=1$ if $b$ was used instead of the reference program, and $Z_b=0$ otherwise.

We trained and evaluated this extended holistic model on our split data-set.
In terms of prediction accuracy, the extended model beats the traditional methodology by almost $20$ percentage points, up to $67.6\%$ on unseen data.
Adding the program to the regression model explains an additional 6.4 percentage of variance over our initial holistic model, nearly 13\% more variance than the ``fuzzer-only'' model.


These improvements indicate that there are still substantial \emph{program-specific} effects that are not captured by the program properties recorded we recorded for this instantiating study.
In other words, the coefficients depicted in  \autoref{fig:mlr} give us a sense of the \emph{net} effect of the corpus properties across programs, but individual effects vary substantially.
For example, overall, Initial Corpus Coverage may have a significant effect on the relative performance of Libfuzzer vs. AFL as found in \textbf{IRQ4}. However, for some programs, this relationship may be somewhat weaker or stronger.
Thus, it is crucial for benchmarking frameworks to maintain a large, representative sample of target programs to avoid systematic bias with respect to program and corpus properties.

This observation further motivates our proposed methodology in future work, as exploratory regression analysis (\textbf{IRQ3}) can identify pertinent benchmark properties and hypothesis testing (\textbf{IRQ4}) can confirm their effects.
Because regression analysis is a sufficiently general technique, it is easy to incorporate additional benchmark properties---numeric, ordinal or categorical---as researchers identify them.
We look forward to future holistic investigations, which examine additional properties to further explain these differences across programs.


\subsubsection{Assumptions}

\begin{figure}
    \centering
    \includegraphics[width=0.45\textwidth]{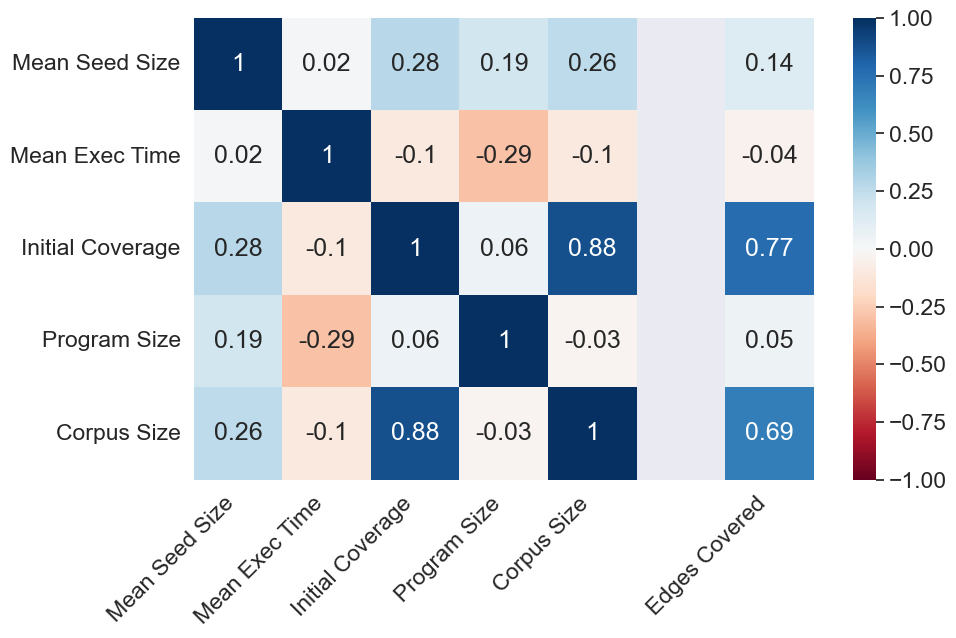}\vspace{-0.2cm}
    \caption{Correlation matrix for Overall Impact Across Benchmarks and Fuzzers (Spearman's $\rho$).}
    \label{fig:correlationmatrix}
    \vspace*{-0.1in}
\end{figure}

To apply multiple linear regression, we require five assumptions to be met:
(i)~a linear association between explanatory and response variables [\emph{linearity}], (ii)~no high correlation among explanatory variables [\emph{no multicolinearity}], (iii)~a constant variance of residuals [\emph{homoscedasticity}], (iv)~normality of the residuals [\emph{normality}] and (v)~independence of observations [\emph{independence}].




\begin{figure}
    \includegraphics[width=0.5\textwidth]{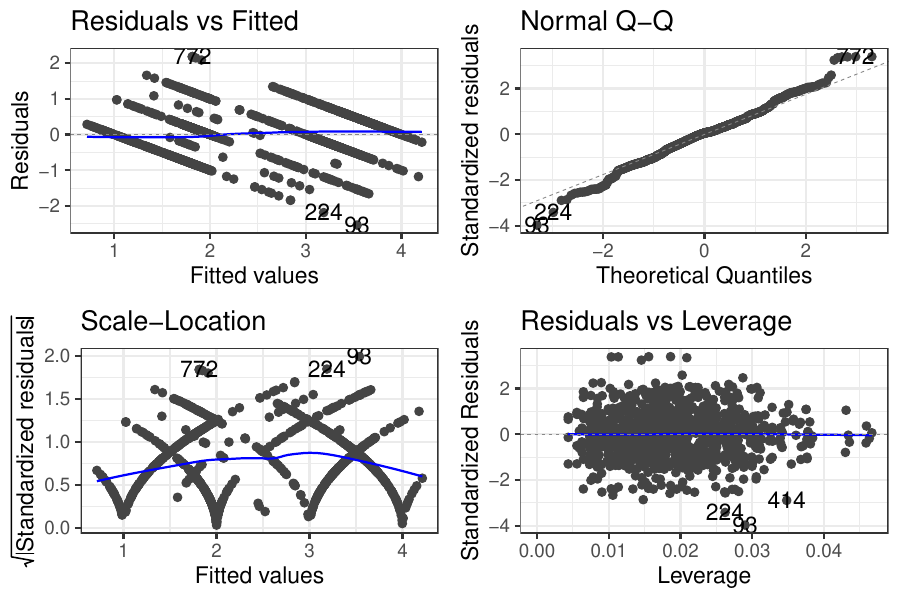}
    \caption{Model Diagnostics}
    \label{fig:mlrdiagnostics}
    \vspace*{-0.1in}
 \end{figure}

\paragraph*{Linearity} For a multiple linear regression model to be accurate, the relationship between the explanatory and response variables must be correctly modeled with a linear formula.
By using the rank transformation, we eliminate monotonic non-linearity, typically observed in fuzzing evaluations \cite{bohme2020expfuzzing}, from the model. 
We observe that all of our transformed predictors appear to be linear with respect to the fuzzer ranking in our supplementary materials (\autoref{sec:data}).

\paragraph*{Multicolinearity} If some explanatory variables are highly correlated, it becomes difficult to interpret the model as the joint effect can be arbitrarily spread among correlated predictors.
In \autoref{fig:correlationmatrix}, we can only see one association that is not weak or negligible, i.e., a very strong correlation between the size and coverage of the initial seed corpus.
As a result, we exclude the corpus size from our model.
For reasons of model parsimony, we also exclude the proportion of equalities, inequalities, and calls to shared library from the multiple linear regression model.
These program properties have no discernible independent impact on individual or relative performance of our fuzzers.

\paragraph*{Normality} To determine statistical significance of regression coefficients analytically, the residual errors of a linear regression model must be approximately normally distributed.
Instead, we obtain confidence intervals \emph{computationally} using a non-parametric bootstrap.

\paragraph*{Homoscedacity} Linear regression models assume that the variance of the residuals is constant over the predicted values of the response variable. We can assess homoscedacity by examining the residual plot or the scale-location plot of the square root of the standardized residuals against the fitted values. In the latter, we should see a flat line if our data is homoscedastic. However, \autoref{fig:mlrdiagnostics} (left) shows that this might not be the case for our data. There seems to be greater variance around the middle ranks, likely due to ties in fuzzer performance (which we break by assigning the average rank).
We therefore use the \emph{wild bootstrap} to reduce the impact of heteroscedasticity on our confidence intervals \cite{wu1986jackknife} \cite{liu1988bootstrap}.
Intuitively, the wild bootstrap repeatedly resamples the residuals of a linear model fitted to the original and scales them by a random variable to give additional data-points without making assumptions that the residuals are similarly distributed.



\paragraph*{Independence} Regression models also assume that the individual samples of data are not dependent on each other.
We use Fuzzbench for our testing setup, which should not introduce any dependence between trials. We validate the independence of the residuals of our data with the Durbin-Watts test \cite{durbin}, obtaining a p-value of $0.794$.





\subsubsection{Variable Selection and Model Parsimony}
\changedminor{When constructing a statistical model for sample data, it is generally desirable to minimize model complexity when attempting to maximize the explanatory power of the model.
Doing so makes the model more interpretable and also reduces the chances of overfitting to the sample data.
Adding variables to a model will typically increase its effectiveness on the sample data, but this comes at the cost of making the model more complex.
Balancing these two requirements is an inherent tradeoff in regression analysis.}

\changedminor{Thus, in constructing our multiple regression model, we picked a \emph{small} subset of properties that (1) we reasonably expected to have an impact on fuzzer effectiveness to demonstrate our approach (2) were easily measurable, (3) that were independent of each other to avoid issues with multicollinearity and overfitting and (4) that were well represented in our sample population of programs integrated with Fuzzbench.
We identify several properties in Section \ref{sec:setuprand} satisfying (1). To ensure (2), we omit other difficult-to-measure properties such as ``seed validity’’~\cite{seedsel21}.
In pursuit of (3), we discarded several properties which were expected or that we found empirically to be highly correlated with others.\footnote{\changedminor{We note that the discarded properties which are strongly correlated with others, such as cyclomatic complexity, may still have value in the model, but a much larger sample size would be necessary to separate out the smaller, individual effects of these variables.}}
Specifically, we excluded the number of memory-unsafe accesses, cyclomatic complexity, and test suite size which are all known or expected to correlate strongly with program size~\cite{shepperd1988critique}. 
We measured other program and corpus properties, including the quantiles of the execution times, quantiles of seed sizes in the initial corpus, as well as the number and types of constraints covered by the initial corpus~\cite{hazimeh2020magma}. However, we found these to be of negligible individual impact and thus excluded them from the final model.
We refer the interested reader to the rich literature on variable selection~\cite{heinze2018variable}.
Finally, we omit other variables --such as whether or not the program has an input ``dictionary''-- due to the small number of representatives in our sample population (4); without a significantly larger number of representatives, we cannot make any statistical claims about the impact of these variables.}

\changedminor{What about important benchmark properties that are not explicitly modeled?
These typically appear as additional unexplained variance in the fitted model (assuming the sample is representative).
Like all statistical methods, our methodology can only account for variables present in the model and observed in the sample data.
If the additional variables were included, the resulting model would have a greater explanatory power, but there would be a corresponding increase in model complexity and thus increased risk of overfitting our sample data.
While there will almost always be \emph{some} unexplained variance after fitting a model, researchers can make incremental improvements and reduce this unexplained variance by running new experiments and measuring new properties.
When they do, they can use our methodology to determine whether these newly measured properties have a significant impact on evaluation outcomes independently of other variables.}

\section{Threats to Validity}\label{sec:threats}

Like other empirical studies, our instantiating studies have several potential threats to validity which we have attempted to mitigate.

\paragraph*{Threats to internal validity} are aspects of our study that may introduce systemic bias.
We minimize experimenter and confirmation bias in our experiment design by utilizing Fuzzbench, an existing benchmarking tool prepared by independent practitioners and researchers, for our experimental setup. 
We reduced the chance of selection bias in our choice of fuzzers by only choosing from state-of-the-art fuzzers with widespread adoption.
Additionally, because we are not attempting to make claims about the performance of any particular fuzzer but rather \textit{potential} covariates in benchmarking runs, impact of selection bias is reduced.
Another possible source of bias was our use of AFL++'s corpus minimization tool to reduce our seed pool for each experiment, which could bias the initial minimized corpora.
We mitigate this risk by only reducing with respect to edge-coverage, which we believe to be an uncontroversial criteria for corpus reduction across grey-box instrumentation tools.
Alternative tools using LLVM coverage data were orders of magnitude slower, to the point that they are not practical for fuzzing practitioners.
Additionally, prior work \cite{seedsel21} has outlined the importance of corpus reduction for fuzzing effectiveness, and thus running experiments without minimized corpora would have threatened the external validity of our study.
Finally, it is important to compare research claims to a baseline to demonstrate that these claims represent an improvement over existing alternatives.
We are not aware of any alternative evaluation frameworks that incorporate and account for covariates in a non-parametric analysis.
However, we do compare with the existing state-of-the-practice for fuzzer evaluations in Section \ref{sec:assumptions}.

\paragraph*{Threats to external validity} could harm the generality of these results beyond the scope of our study.
\changed{For our two controlled experiments, we only seek to demonstrate that certain specific properties \emph{can} influence the outcomes of benchmarking runs.
While we chose hypotheses for these experiments that we reasonably believe may generalize, we cannot say that our results for \textbf{IRQ1} and \textbf{IRQ2} will hold for other fuzzers or subject programs without additional experiments.}
Our holistic methodology in \autoref{sec:holistic} describes how to account for these and other properties in practice on arbitrary other programs.
For the third instantiating experiment in \autoref{sec:t3}, we attempt to minimize this risk with respect to our benchmark programs by utilizing the Fuzzbench benchmark suite, which is broadly representative of open source software available to the fuzzing community \cite{metzman2021fuzzbench}.
\changedminor{There is a risk that by using an existing fuzzer benchmarking suite, our sample population \emph{only} includes projects which have been previously subjected to heavy fuzzing; this may not be representative of general software programs, many of which have not been fuzzed.
However, creating a new benchmark with programs that have never been fuzzed would greatly increase the cost of our study and risk introducing other biases in the benchmark (i.e. there may be reasons \emph{why} these projects have not been fuzzed -- e.g. they are very small or not widely used).
As such, we opt to use a well-established benchmark for our initial study, with the hope that this methodology can be applied to other benchmarks and programs in future research.}
Additionally, we follow the recommendation of B\"ohme et al.~\cite{benchmarking}, that $\geq$10 benchmark programs should be sufficient for coverage-based benchmarking, spending over 1000 CPU-hours gathering data for our analysis.
We also make our data and experimental setup available for replication.
We attempt to minimize external validity risk with respect to our fuzzers by choosing what we believe to be a representative set of state-of-the-art fuzzers.
AFL and LibFuzzer are the baseline implementations for 10/11 state-of-the-art fuzzers supported in Fuzzbench's main experiment.
AFL++ and Entropic are among the latest and highest performing variants of our two baseline implementations in recent benchmarking runs.
Similarly, SymSan and Eclipser are both recently published tools at top research venues, with promising experimental results.

\paragraph*{Threats to construct validity} concern whether or not the data collected in our study measure what we claim.
\changed{The execution speed parameter assessed in \textbf{IRQ1} must be interpreted with caution.
We fixed all other variables for this experiment, but in many real-world programs, execution time is correlated with other variables.
Our results may reasonably hold when other variables such as the number of instructions remain constant as execution time increases (e.g. if there is increased blocking on I/O resources).}

In this study, we are examining the impact of various variables on ``fuzzer effectiveness''.
The measure we used, code-coverage, has long been used as proxy for fuzzer performance,
but there is a risk that our results do not translate to bug-finding effectiveness. 
However \cite{benchmarking} found that coverage based benchmarking is correlated with bug finding ability.
Additionally, because of the sparsity of bugs, using even the largest bug-based benchmarks available (around 200 bugs) as our metric for fuzzer effectiveness would threaten external validity of our study.

\paragraph*{Threats to conclusion validity}
may lead to misinterpretations of our data.
We mitigate the risk by checking the assumptions of each technique used.
In the case of our regression models, we also use the non-parametric wild bootstrap \cite{wu1986jackknife} to obtain confidence intervals, rather than analytical methods dependent on the distribution of the residuals.

%
Another potential threat to conclusion validity is our use of the rank transformation.
We did so to reduce the impact of outliers, scale the data for each program such that it is comparable, and eliminate the impact of any monotonic non-linearity on our regression analysis.
Indeed, without the rank transformation, our data appears to be non-linear with no obvious higher order pattern, in addition to having several strong outliers, and thus would not be amenable to regression analysis.
While the rank transformation is a common statistical technique for non-parametric analysis \cite{kruskalwalis,iman1979use,mannwhitney,nonparametrics,conover1981rank},
this transformation introduces a layer of indirection to our analysis; where we 
report correlations and trends in the ranked data, and not the underlying raw data.

Finally, testing multiple hypotheses can increase the probability of Type I error, known as the issue of multiplicity.
While our holistic evaluation methodology can provide confidence intervals for an arbitrary number of effects, we advocate only formally testing a single or small number of hypothesis per experiment with our holistic methodology, as in \ref{sec:hyp}.
However, any time multiple tests are conducted, it is important to at least \emph{consider} adjusting for multiplicity.
Unfortunately, most adjustments typically decrease statistical power, often substantially so, and thus should not be applied blindly \cite{nakagawa2004farewell, perneger1998s}.
As such, whether to conduct such a correction depends on the context and whether false negatives or false positives are more important to avoid.
For academic research, usually false positives are less desirable than false negatives, so we use the Bonferroni correction for claims of significance in experiments for \textbf{IRQ 1} and \textbf{IRQ 2}.
Note that because of our experimental setup, the loss in statistical power for these experiments is of minimal impact.
For \textbf{IRQ 4}, we adjust the alphas for each regression coefficient's confidence interval using the Holm–Bonferroni~\cite{holm1979simple} method, which is strictly more powerful than the standard Bonferroni correction.

\vspace{-0.1cm}
\section{Related Work}
In disciplines of computer science which often utilize empirical evaluations, the impact of the evaluation setup on its outcome has previously been noted, but---to the best of our knowledge---not systematically studied. We could not find other methodologies that \emph{account} for or quantify the impact of the benchmark on the benchmarking outcome \emph{during} the evaluation.
For instance, Kudela recently found that 47 of 90 evolutionary algorithms exhibit a strong bias towards the center of the search space due to prominent benchmarks in that field having solutions at or near the center \cite{kudela2022critical}.
In machine learning, Japkowicz \cite{japkowicz2006question} warns that focus on performance measures in research might obscure important behaviors of the algorithms under consideration.
In computer architecture, Panda et al. \cite{panda2018wait} study the characteristics of the SPEC2017 CPU benchmark suite primarily to identify a \emph{subset} of benchmark programs and inputs that can approximate the results of the \emph{entire} benchmark suite.
These works all acknowledge an impact of the choice of benchmark on the outcome of the evaluation, but provide no guidance to account for these effects in practice.

Similarly, in automated software testing, Herrera et al. \cite{seedsel21} found that the properties of the seed corpus had a substantial impact on the performance of a fuzzer.
For example, they found that running fuzzers on the \texttt{readelf} program using i.e. an empty corpus, a single valid ELF file, and a large corpus of valid ELF files gave completely different results.
This is an example of a \emph{controlled} experiment where one variable, the corpus size, is changed while others are held constant.
However, the control Herrera et al. exercise in their experiments is incomplete -- the corpus size may be correlated or even confounding other important variables which cannot be held constant.
Indeed, in the third instantiating study of our framework, we find that corpus size and the initial coverage of the corpus are highly correlated, both confirming and shedding additional light on their experiments.
As a key takeaway from their work, Herrera et al. recommend that evaluators vary corpora to see how they impact the fuzzers being assessed.
However, they not provide a methodology to investigate properties other than corpus size.
Our work gives a concrete workflow that evaluators can use for arbitrary properties of the corpus or benchmark to make this recommendation actionable.

In terms of guidelines for the evaluation of automatic software testing tools (i.e., fuzzers), most recommendations are concerned with the sound statistical analysis of effect size and statistical significance.
Arcuri and Briand \cite{arcuri2011practical} \cite{arcuri2014hitchhiker} provide an in-depth discussion of statistical, non-parametric measures that are particularly applicable in the context of automatic software testing. 
Klees et al. \cite{klees2018evaluating} later drew on these guidelines to identify specific issues with experimental design for fuzzer evaluations, such as lack of repetition, too few target programs, and lack of consideration for the impact of initial corpora.
However, none of the previous guidelines provide clear steps for an evaluation methodology that incorporates these covariates, as presented in this paper.


Several works have emerged establishing benchmarks for fuzzing tools both generally, as in the case of Fuzzbench \cite{metzman2021fuzzbench} and for specific classes of applications such ProFuzzBench \cite{profuzzbench}.  UNIFUZZ \cite{li2021unifuzz} attempts to provide a comprehensive evaluation platform by studying the impact of fuzzer-specific factors like instrumentation methods and crash analysis tools. However it does not study the impact of benchmark properties. In fact, like other fuzzing benchmarks, UNIFUZZ uses a static seed set for each benchmark program. As a result it does not study the differences in fuzzer evaluation results that arise from the choice of benchmark properties. Magma \cite{hazimeh2020magma} focuses on a suite of real-world bugs.
Other projects have gathered data-sets of real-world bugs for the evaluation of various testing and debugging strategies \cite{just2014defects4j} \cite{gyimesi2019bugjs}.
All of these works use methodology similar to  Klees et al. \cite{klees2018evaluating} or Arcuri and Briand \cite{arcuri2014hitchhiker}, thereby, neglecting the influence of benchmark properties on testing outcomes.

Synthetic benchmarks \cite{gavitt16lava} \cite{patra21semseed} \cite{pewny16evilcoder} \cite{Subhajit2018Apocalypse} provide an alternative path for more holistic benchmarking.
For example, Zhu et al. \cite{zhu2019feature}, propose creating artificial target programs with large numbers of features known or suspected to be problematic for fuzzers.
They compare relative performance of two fuzzers on one such program in terms of artificial ``bugs'' found, observing a difference in behavior between AFL and AFLFast \cite{aflfast}.
This approach, however, cannot account for different effects from the seed corpora for such artifical programs.
Additionally,
there remain doubts as to whether results from such synthetic benchmarks are representative of behavior in typical real-world programs \cite{bundt21evalsynthetic} \cite{Geng2020empiricalartificialvuln} \cite{boehme2014corebench}.
Recent work by Lyu et al. \cite{lyu2022slime} suggests  program senstive energy allocation procedure to determine the power schedule of a fuzzer. 


While our study focused on coverage as the primary metric for fuzzer performance, we expect program and corpus properties to have impacts beyond increased coverage.
Another metric commonly used to evaluate testing tools is mutant injection \cite{papadakis2019mutation}.
The ground truth for testing effectiveness is bug-finding ability. 
Inozemtseva et al. \cite{inozemtseva2014coverage} first investigated the correlation between code coverage and bugs found by Java unit testing suites.
B{\"o}hme et al. \cite{benchmarking} more recently found coverage to be a good objective function for fuzzers, given the sparsity of real-world bugs.
We believe there are future research opportunities in assessing the impact of covariates with respect to these alternative metrics. 


\paragraph{Ensemble Fuzzing}
\changed{Ensemble fuzzing (a.k.a. collaborative fuzzing or fuzzer composition) is a related subfield of research.
The goal of an ensemble is to select the best performing set of fuzzer(s) for a given workload. 
Our multiple-linear regression model could, in theory, be used to predict a set of fuzzers that perform best for a workload (indeed we use prediction accuracy to validate our model in Section \ref{sec:assumptions}), but this is not the primary purpose of our framework.
We aim to provide developers and researchers with the tools to conduct nuanced evaluations of fuzzer performance; in essence to understand which variables contribute to a particular fuzzer performing best on a given benchmark, and account for the impacts of those variables.
Additionally, existing ensemble fuzzers are more focused on \emph{characteristics of the fuzzers} themselves rather than the properties of benchmark programs discussed in this work.
For example, \texttt{autofz}~\cite{fu2023autofz} dynamically picks the fuzzer(s) making the best progress during an ongoing fuzz campaign, but cannot explain why those particular fuzzers are performing well (or why other fuzzers are not) on that particular workload.
The \texttt{EnFuzz}~\cite{chen2019enfuzz} authors compose fuzzers looking for diversity in their broad characteristics, such as seed selection strategy or mutation strategy.
Similarly, \texttt{Cupid}~\cite{guler2020cupid} empirically assesses which fuzzer pairs are most complementary in terms of code-coverage.
One possible application of our methodology could be to shed further light into which program or corpus properties each individual fuzzer is leveraging in a successful ensemble.}

\section{Perspective: Beyond Fuzzing}

Benchmarking allows researchers and practitioners to make empirical claims about the properties of a newly proposed technology---fuzzers included. \changed{In this paper, we study how to assess the degree to which the outcome of an evaluation depends on the specific properties of the benchmark that is used for the evaluation.}
In particular, we propose using control and randomization to assess the effects of benchmark properties on the evaluation outcome. 
Our approach can be used to study arbitrary variable effects, both in isolation and in combination with each other.
While we describe our approach in the context of fuzzing, this methodology can be applied to other evaluations in other domains.

Within our domain of expertise, we showcase three instantiating studies leveraging this methodology, finding several examples of statistically significant effects not accounted for in current evaluations.
Fuzzing benchmarks should be diverse and randomized with respect to properties like program execution time, program size, seed origin, and initial coverage.
Furthermore, we hope researchers will use holistic methods to identify additional properties of benchmarks which impact evaluation outcomes.
Benchmarks such as Fuzzbench can incorporate our changes, randomizing initial corpora and other parameters to provide more robust outcomes.
Finally, fuzzer developers can adopt our methodology to account for previously identified covariates in their evaluations.

\vspace{-0.1cm}
\section{Data Availability}
\label{sec:data}

We include our infrastructure, data, and analysis at:
\begin{quote}\centering
\small{\textbf{\textcolor{blue}{
\url{https://figshare.com/s/de961f6206f786997e87}
}}} \end{quote}
\vspace{0.1cm}
We will make these artifacts publicly available upon acceptance.

\section{Acknowledgements}
This research is supported by the National Research Foundation, Singapore, and Cyber Security Agency of Singapore under its National Cybersecurity R\&D Programme (Fuzz Testing $<$NRF-NCR25-Fuzz-0001$>$).
Any opinions, findings and conclusions, or recommendations expressed in this material are those of the author(s) and do not reflect the views of National Research Foundation, Singapore, and Cyber Security Agency of Singapore. 
This research is also partially funded by the European Union. Views and opinions expressed are however those of the author(s) only and do not necessarily reflect those of the European Union or the European Research Council Executive Agency. Neither the European Union nor the granting authority can be held responsible
for them. This work is supported by ERC grant (Project AT\_SCALE, 101179366).

\bibliographystyle{ACM-Reference-Format}
\bibliography{bibtex}

\end{document}